\documentclass{cjaa}                    

\usepackage{graphicx}                   
\input{epsf.sty}                        
\input{psfig.sty}                       

\begin{document}

   \title{GRAPE - A Balloon-Borne Gamma-Ray Polarimeter Experiment}
   


   \author{P. F. Bloser
      \inst{1}\mailto{}
   \and J. S. Legere
      \inst{1}
   \and J. R. Macri
      \inst{1}
   \and M. L. McConnell
      \inst{1} 
   \and T. Narita
      \inst{2} \\
   \and J. M. Ryan
      \inst{1}
      }
   \offprints{P. F. Bloser}                   

   \institute{Space Science Center, University of New Hampshire, Durham, NH  03824\\
             \email{Peter.Bloser@unh.edu}
        \and
	Department of Physics, College of the Holy Cross, Worcester, MA  01610
          }

   \date{Received~~2001 month day; accepted~~2001~~month day}

   \abstract{
This paper reviews the development status of GRAPE (the Gamma-Ray Polarimeter Experiment), a hard X-ray Compton Polarimeter. The purpose of GRAPE is to measure the polarization of hard X-rays in the 50-300 keV energy range. We are particularly interested in X-rays that are emitted from solar flares and gamma-ray bursts (GRBs), although GRAPE could also be employed in the study of other astrophysical sources. Accurately measuring the polarization of the emitted radiation will lead to a better understating of both emission mechanisms and source geometries. The GRAPE design consists of an array of plastic scintillators surrounding a central high-Z crystal scintillator. The azimuthal distribution of photon scatters from the plastic array into the central calorimeter provides a measure of the polarization fraction and polarization angle of the incident radiation.  The design of the detector provides sensitivity over  a large field-of-view ($>\pi$ steradian).  The design facilitates the fabrication of large area arrays with minimal deadspace.   This paper presents the latest design concept and the most recent results from laboratory tests of a GRAPE science model.
   }

   \authorrunning{P. F. Bloser et al. }            
   \titlerunning{Balloon-Borne Polarimeter }  


   \maketitle
%
%
\section{Introduction}           
\label{sect:intro}

There are four properties of the radiation emitted from cosmic sources that can be measured: energy, intensity, direction, and polarization. Many experiments have been conducted to measure the energy, intensity and direction of these incoming photons, but measurements of polarization have been lacking largely due to the lack of instrumentation with sufficient sensitivity. Polarization measurements have become a powerful tool for astronomers throughout the electromagnetic spectrum. It is believed that by accurately measuring polarization levels from solar flares and GRBs we will be able to better understand both the emission mechanisms and source geometries producing the observed radiation (e.g., Lei et al. 1997).
Here we report on  the development status of GRAPE (the Gamma-Ray Polarimeter Experiment), a hard X-ray Compton Polarimeter designed to measure the polarization of hard X-rays in the 50-300 keV energy range.  As described here, the GRAPE design is most suitable for studies of either gamma-ray bursts or solar flares, as part of a long-duration balloon platform or as part of a satellite platform, although it could easily be adapted into the design of an imaging polarimeter.


\section{Principal of Operation}
\label{sect:data}

Measuring the polarization of X-rays in the 50-300 keV energy range is most easily done by taking advantage of the Compton scattering process.  In Compton scattering,  photons tend to be scattered at a right angle with respect to the incident electric field vector. In the case of an unpolarized beam of incident photons, there will be no net positive electric field vector and therefore no preferred azimuthal scattering angle; the azimuthal distribution of the scattered photons will be uniform.  However, in the polarized case, the incident photons will exhibit a net positive electric field vector and the azimuthal  distribution will be asymmetric.  

The ultimate goal of a Compton scatter polarimeter is to measure the azimuthal distribution of the scattered photons. In general, a Compton scatter polarimeter consists of two detectors to determine the energies of both the scattered photon and the scattered electron. One detector, the scattering detector, provides the medium for the Compton interaction to take place. This detector must be designed to maximize the probability of there being a single Compton interaction with a subsequent escape of the scattered photon.  This requires a low-Z material in order to minimize photoelectric interactions. The area of the scattering detector which is exposed to the photon beam is also an important factor in determining the effective area of the polarimeter. The primary purpose of the second detector, the calorimeter, is to absorb the full energy of the scattered photon.

The relative placement of the two detectors defines the scattering geometry. For incident photon energies below 100 keV, the azimuthal modulation of the scattered photons is maximized if the two detectors are placed at a right angle relative to the incident photon beam (at a Compton scatter angle of $\theta = 90^{\circ}$). The positioning of the two detectors must also be arranged relatively close to each other so that the second detector subtends a sufficiently large solid angle to achieve the required detection efficiency. At the same time, a larger separation between the detectors provides more precise scattering geometry information. The accuracy with which the scattering geometry can be measured determines the ability to define the modulation pattern (Fig. 1) and therefore has a direct impact on the polarization sensitivity. Here, one must compromise between total efficiency (small detector separation) and the ability to define the modulation pattern (large detector separation).

\begin{figure}
	\centering
	\includegraphics[width=3.50in]{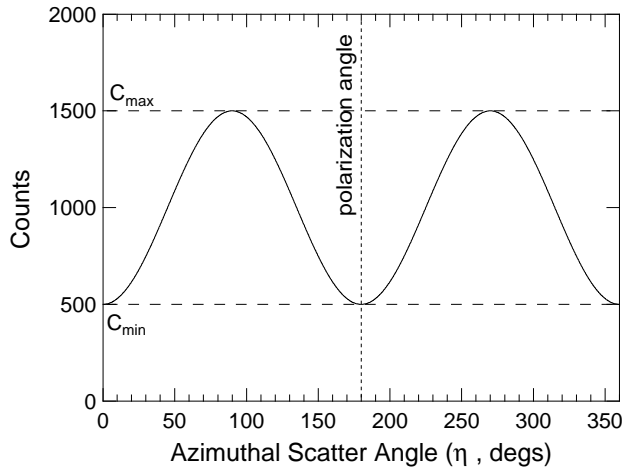}
	\caption{The azimuthal scatter angle distribution provides a measure of both the level of polarization (based on the values of $C_{min}$ and $C_{max}$) and the polarization angle (based on the location of the minimum in the distribution).}
	\label{fig_sim}
\end{figure}

In principle, the scattering detector need not be active.  At soft X-ray energies ($<10$ keV), for example, the scattering process results in a very small energy loss that can be difficult to detect.   For polarimeters designed to measure soft X-rays, the scattering element is usually passive, its only requirement being that it be designed to maximize the probability of a single photon scatter.  At higher energies, an active element is certainly preferred as a means to minimize background, but it is certainly not required.

The development and design of the GRAPE detector has evolved through three science models (McConnell et al. 1998, 1999a, 1999b, 1999c, 2000, 2002a, 2003, 2004; Legere et al. 2005). Each one represented a successive improvement, but all three essentially operate under the same underlying principle of operation. A high-Z material, the calorimeter, is surrounded by multiple plastic scintillation detectors that serve as a target for the Compton scattering. The entire array is contained within a single light-tight housing. The plastic scintillators are made of a low-Z material that maximizes the probability of a Compton interaction. Ideally, photons that are incident on the plastic scintillator array will Compton scatter only once, and then be absorbed by the calorimeter. For such an event we measure the energy of the scattered electron in the plastic and the deposited energy of the scattered photon in the calorimeter. With multiple plastic scintillators surrounding the calorimeter, we can determine the azimuthal scatter angle of each valid event. A histogram of these data represents the azimuthal modulation pattern of the scattered photons, which provides a measurement of the polarization parameters (magnitude of the polarization and polarization angle) of the incident flux.

In order to accurately measure the azimuthal modulation (and hence the polarization parameters), we need to correct for geometric effects specific to the individual detector design. When the azimuthal modulation profile is generated, the distribution not only includes the intrinsic modulation pattern due to the Compton scattering process, but it also includes various geometric effects. One of these effects originates from the specific layout of the detector elements within the polarimeter and the associated quantization of possible scatter angles. Other effects include such things as the nonuniform detection efficiency of the PMT used for detector readout. In order to correctly analyze the data we first measure unpolarized radiation with the polarimeter. These data provide a set of correction factors that incorporate the various geometric effects and that can be used to correct the source data of interest.  In the laboratory, these systematic effects are easy to measure using data from an unpolarized beam. In practice, simulations may be used to determine the unpolarized modulation pattern.

To determine the polarization level of the incident radiation we need to know the azimuthal modulation for a completely polarized beam. We have used simulations based on MGEANT (incorporating the GLEPS polarization code; McConnell et al., 2005, in preparation) to model the response of the polarimeter to 100\% polarized incident radiation. For the laboratory measurements presented below, the simulations include all important components of the lab setup.

\section{Latest Laboratory Results}

Our latest laboratory measurements are based on the third GRAPE science model (SM3).  This compact design is based on the use of flat-panel MAPMT (Hamamatsu H8500; e.g., Pani et al. 2004).  The H8500 is a MAPMT with an array of $8 \times 8$ independent anodes. The 5 mm anodes are arranged with a pitch of 6 mm, occupying a total area of 52 mm $\times$ 52 mm. With a depth of 28 mm, the flat-panel PMT offers a very compact design.

In the latest laboratory tests, the MAPMT was fully populated with 60 plastic scintillators, each measuring 5 mm $\times$ 5 mm $\times$ 50 mm.  Each plastic scintillator element is wrapped in Tyvek{\textregistered} and secured with Kapton{\textregistered} tape to provide optical isolation of each element. An optically-isolated CsI(NaI) calorimeter element, measuring 10 mm $\times$ 10 mm $\times$ 50 mm, is placed at the center of the array.  The whole array assembly is held together with cookie spacers made from Delrin{\textregistered} and coupled to the face of the MAPMT with clear GE silicone.  The silicon offers the advantages of increasing the light output measured in the MAPMT anodes and providing a more rugged assembly.  The completed array is shown in Fig. 2.  During operation, the whole detector is housed within a light-tight aluminum enclosure.  The logic diagram of the detector module is shown in Fig. 3.  For lab testing, we use a $^{137}$Cs calibration source, whose photons are scattered at 90¡ in a block of plastic scintillator.  The scattering produces a beam of photons with a net polarization of $\sim55-60$\%.  The use of plastic scintillator as a scattering medium also provides a relatively efficient means for electronically tagging the scattered photons.  This signal is used in coincidence with the two signals in the polarimeter to identify valid events via a three-way coincidence.

\begin{figure}
	\begin{minipage}[t]{.35\linewidth}
	\centering
	\includegraphics[width=1.75in]{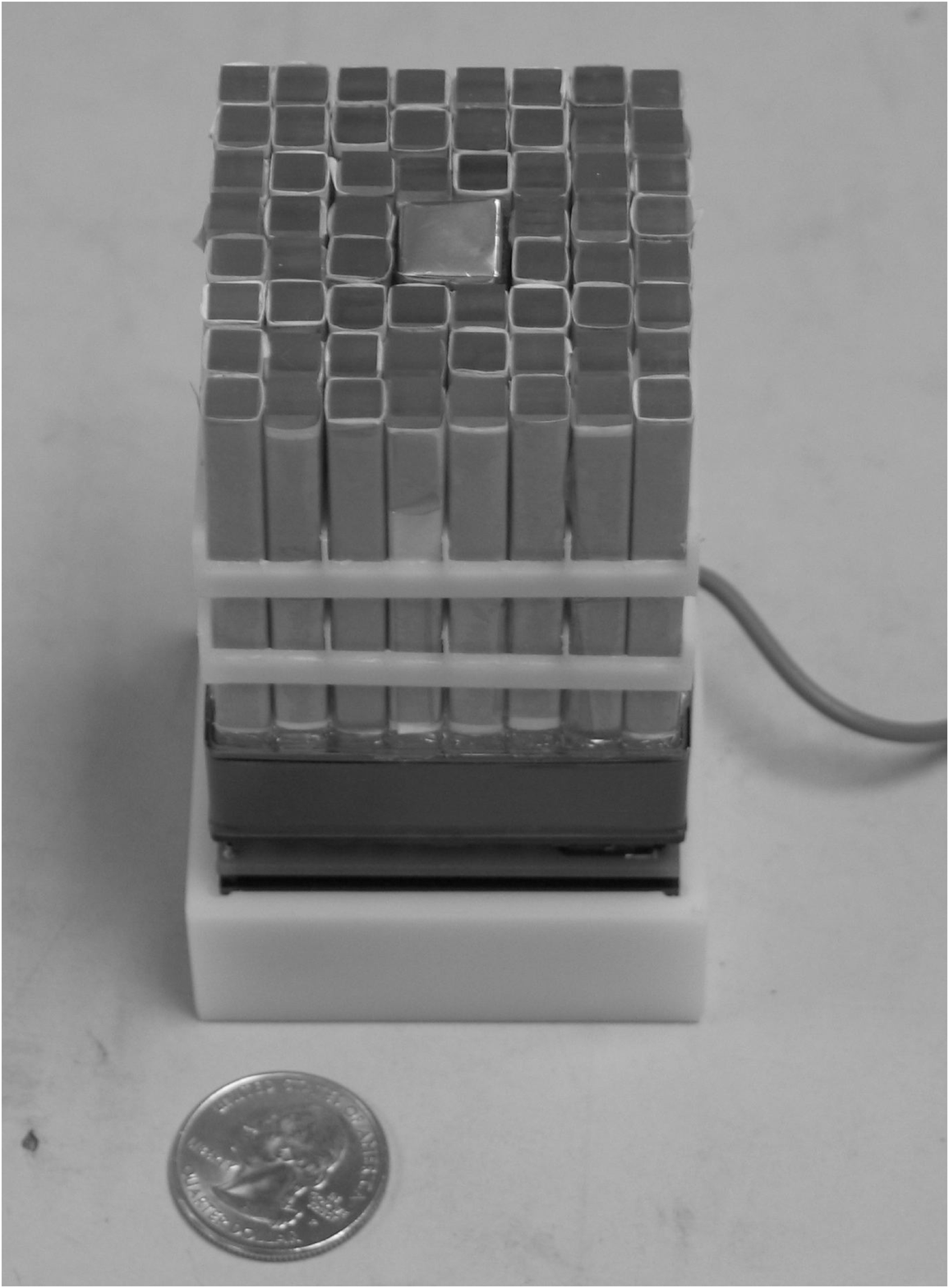}
	\caption{The array of scintillators on top of the flat-panel PMT ready for final assembly.}
	\label{fig_sim}
	\end{minipage}\hfill
	\begin{minipage}[t]{.65\linewidth}
	\centering
	\includegraphics[width=3.00in]{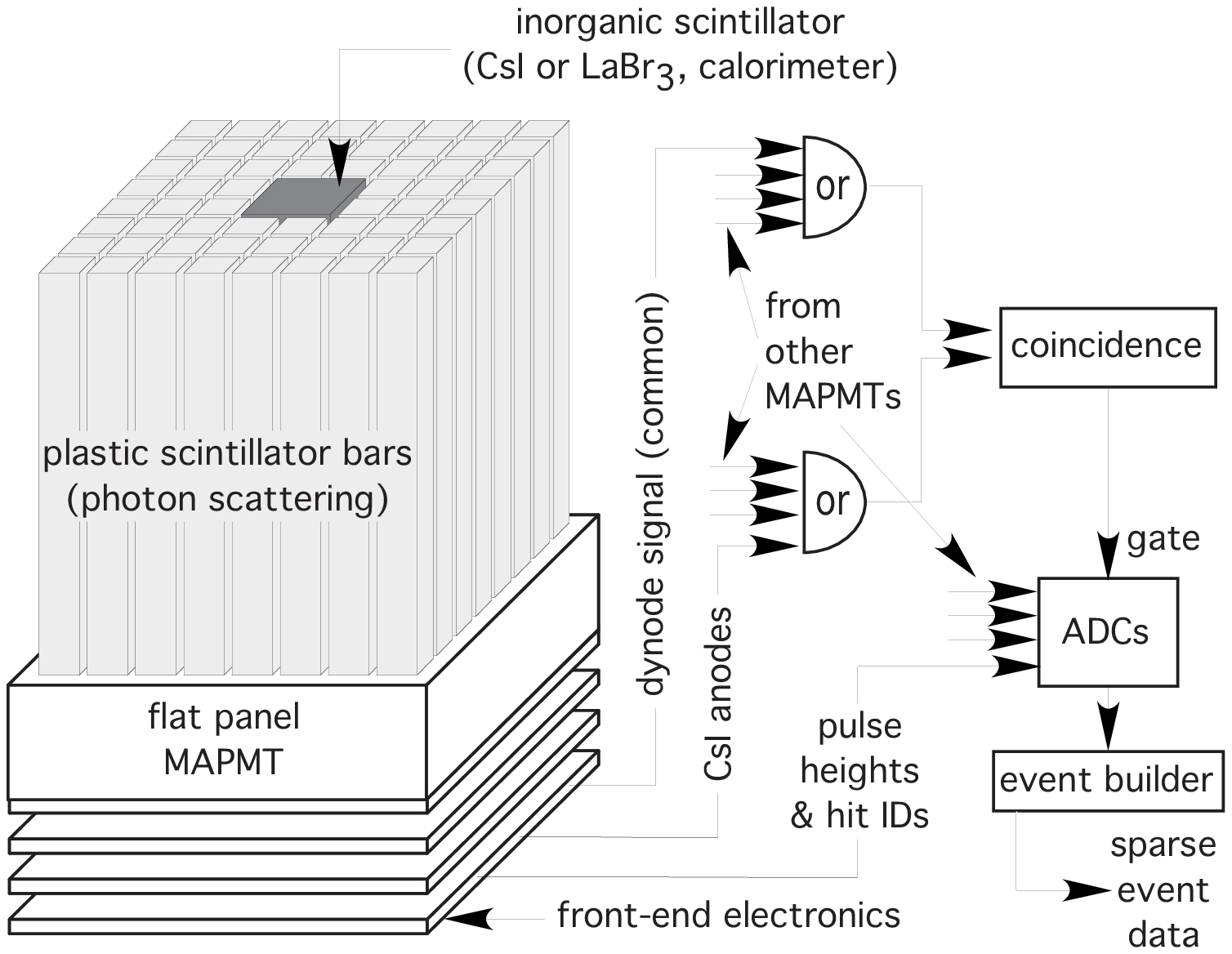}
	\caption{The logic diagram for a single GRAPE module.}
	\label{fig_sim}
	\end{minipage}
\end{figure}

The initial laboratory results with  SM3 proved unsatisfactory. It was discovered that there were light cross-talk issues between the calorimeter and the twelve surrounding plastic elements. We believe that this can be attributed to the close proximity between the calorimeter and adjacent plastics. With a light output of $\sim4$ times that of a plastic scintillator, the CsI(NaI) calorimeter was contributing to the light recorded in the adjacent channels. This virtually eliminated our ability to reconstruct accurate scatter angle distributions.  As we explored possible options for minimizing this crosstalk,  we decided to eliminate the inner twelve plastic scintillator elements from the analysis. This leaves 48 plastic elements in the array for analysis of the azimuthal scatter angle distribution.   The results shown in Fig. 4 and Fig. 5 represent the data collected with innermost plastic elements physically in place, but events occurring in those elements eliminated from the analysis. Data was collected for $0^{\circ}$ and $90^{\circ}$ polarization angles.  The data  yielded measured polarization levels of 56($\pm$9)\% and 55($\pm$7)\%, respectively.  The measured polarization angles of $2^{\circ}(\pm4^{\circ})$ and $90^{\circ}(\pm3^{\circ})$, respectively, are consistent with the laboratory setup.

\begin{figure}
	\begin{minipage}[t]{.46\linewidth}
	\centering
	\includegraphics[width=2.50in]{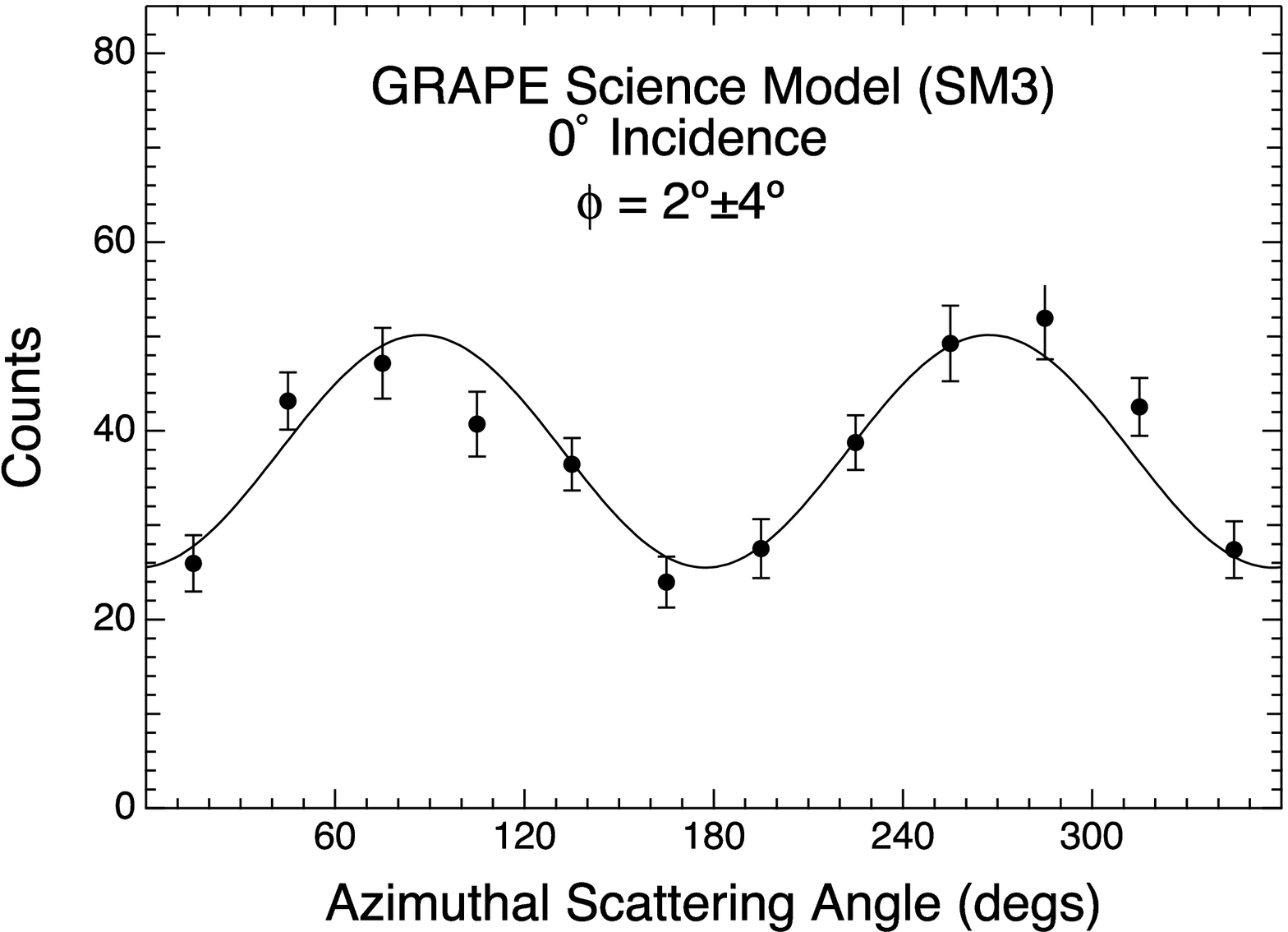}
	\caption{The photon scatter angle distribution measured with GRAPE for an incident polarization angle of $0^{\circ}$.  The measured polarization level in this case was 56\%.}
	\label{fig_sim}
	\end{minipage}\hfill
	\begin{minipage}[t]{.46\linewidth}
	\centering
	\includegraphics[width=2.50in]{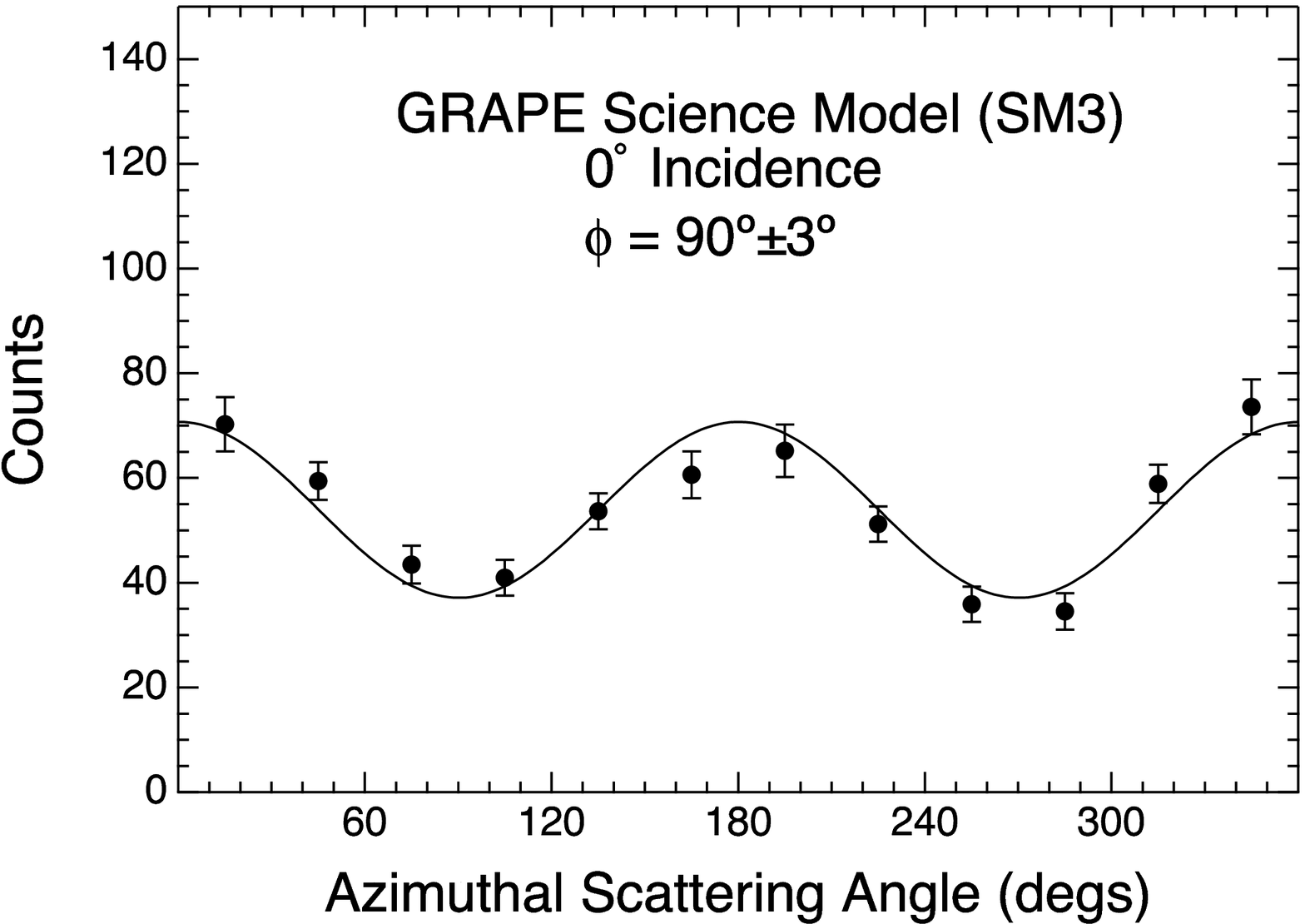}
	\caption{The photon scatter angle distribution measured with GRAPE for an incident polarization angle of $90^{\circ}$.  The measured polarization level in this case was 55\%.}
	\label{fig_sim}
	\end{minipage}
\end{figure}

\section{Future Plans}

The next step for GRAPE is improved coincidence timing and energy resolution. We plan to accomplish this by replacing the central CsI calorimeter with one based on Lanthanum Bromide (LaBr$_3$; van Loef et al. 2001; Shah et al. 2003). This relatively new inorganic scintillator provides an energy resolution that is more than twice as good as NaI at 662 keV (3\% vs. 7\%). A similar improvement in the energy resolution of GRAPE can be expected (Fig. 6). The LaBr$_3$ also provides for better timing characteristics. With decay times of $\sim25$ nsec, it is comparable to the plastic scintillators used and will greatly improve the coincidence timing. 

In order to provide adequate sensitivity, any realistic application of the GRAPE design would involve an array of polarimeter modules, as depicted in Fig. 7 (McConnell et al. 2003). One possible deployment option for a GRAPE polarimeter array would be as the primary instrument on an Ultra-Long Duration Balloon (ULDB) payload.  The ULDB technology currently under development by NASA is expected to provide balloon flight durations of up to $\sim100$ days. A 1 m$^2$ array of GRAPE modules would easily fit within the envelope of a ULDB payload. The ideal configuration for GRB or solar flare studies would be an array that remains pointed in the vertical direction (i.e., towards the zenith) at all times.  In this case, there would be no pointing requirements, only a moderate level of aspect information (continuous knowledge of the azimuthal orientation to $\sim0.5^{\circ}$). An imaging polarimeter could also be designed to match the payload limitations of a ULDB, although the pointing requirements would be much more severe ($<1^{\circ}$ in both azimuth and zenith).  A second deployment option would be as part of a spacecraft payload.  This could be either a free-flying satellite or an add-on experiment for the International Space Station (ISS).

\begin{figure}
	\begin{minipage}[t]{.46\linewidth}
	\centering
	\includegraphics[width=2.5in]{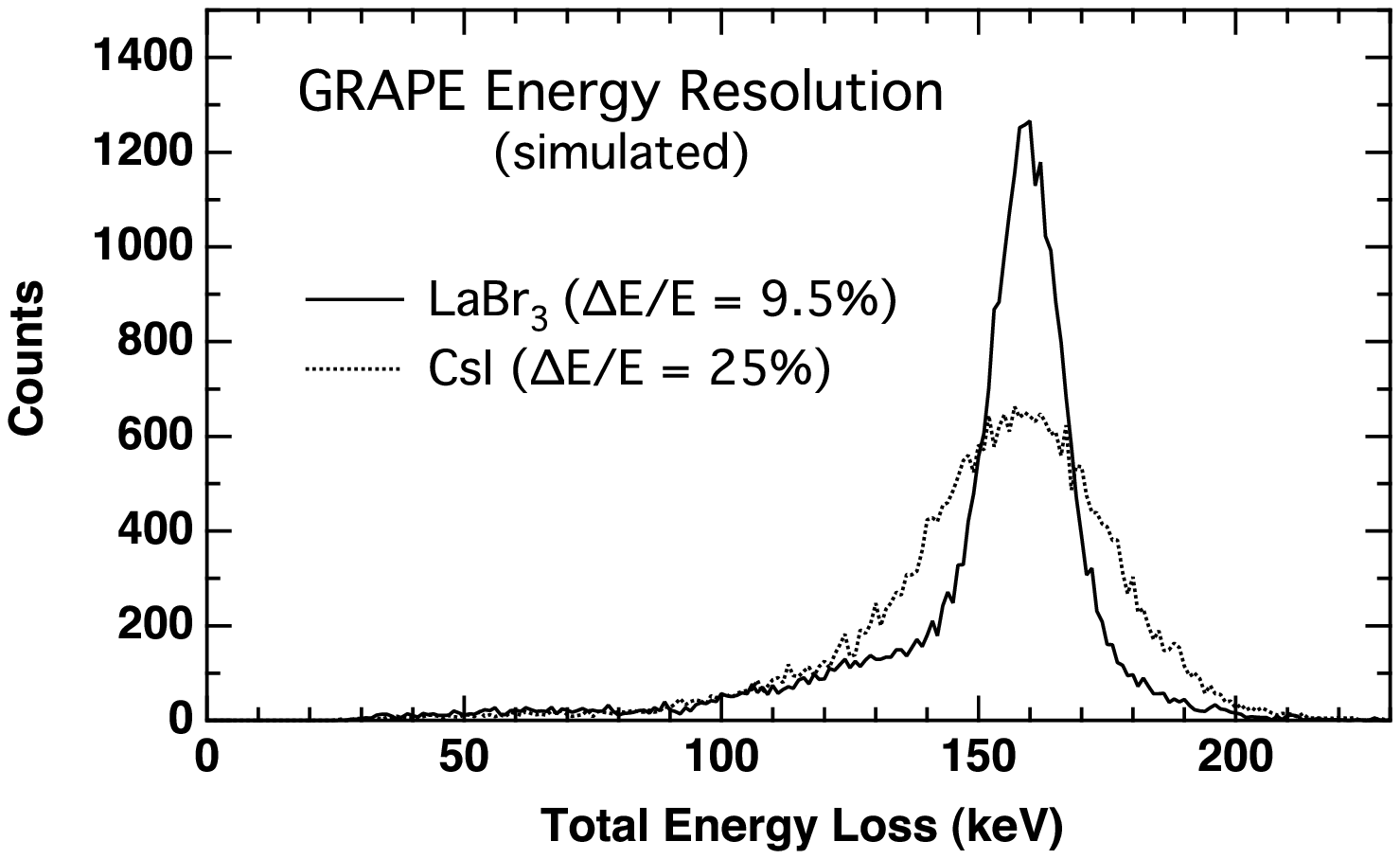}
	\caption{The use of LaBr$_3$ scintillator as a calorimeter would provide a significant improvement in the total energy resolution.}
	\label{fig_sim}
	\end{minipage}\hfill
	\begin{minipage}[t]{.46\linewidth}
	\centering
\includegraphics[width=2.25in]{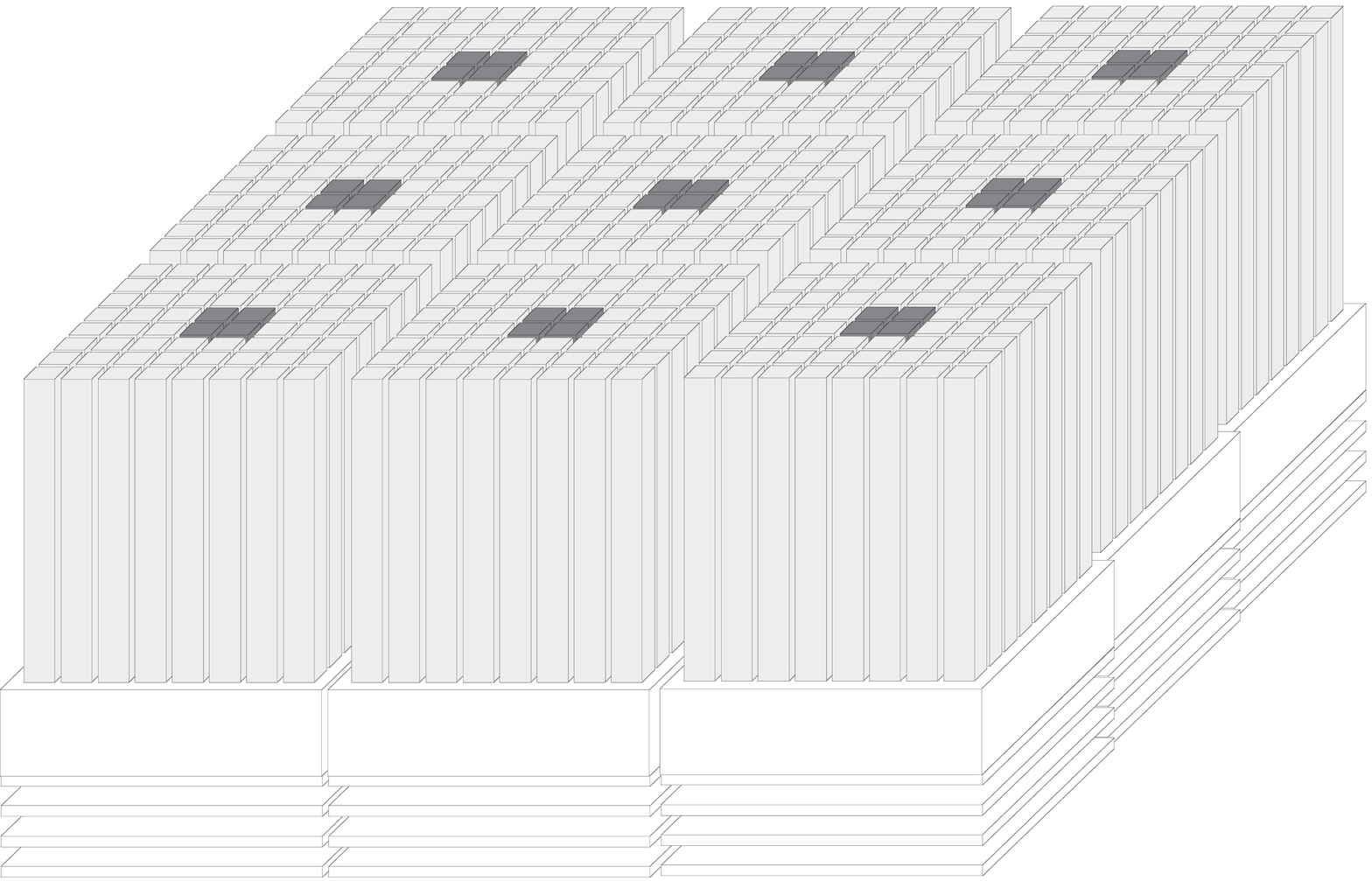}
	\caption{An array of GRAPE modules could be made arbitrarily large.  In practice, such an array would likely be surrounded by an anticoincidence shield. }
	\label{fig_sim}
	\end{minipage}
\end{figure}

\begin{acknowledgements}
This work is currently supported by NASA grants NNG04GB83G and NNG04WC16G. We would like to thank Mark Widholm and Paul Vachon for their support with the MAPMT design electronics. We would also like to thank the Laboratory for Advanced Instrumentation Research at Embry Riddle, Sparrow Corp., Drew Weisenberger, and Sergio Brambilla for their support with the VME data acquisition setup.
\end{acknowledgements}

\bigskip
\noindent
{\b DISCUSSION}
\bigskip

\noindent
{\b REBECCHI:} Where will GRAPE be launched?
\bigskip

\noindent
{BLOSER:} The initial engineering test flight will likely be flown (as a piggyback payload) from Texas or New Mexico.  Eventually, we will want to fly on an Ultra Long Duration Balloon.
\bigskip

\label{lastpage}

\end{document}